\documentclass[aps,pra,twocolumn,superscriptaddress,showpacs]{revtex4}
\usepackage{graphicx}
\usepackage{amsmath}
\usepackage{epsfig}
\usepackage{bm}

\newcommand{\beq}{\begin{eqnarray}}
\newcommand{\eeq}{\end{eqnarray}}

\begin{document}

\title{Increase of entanglement by local $\mathcal{PT}$-symmetric operations}
\author{Shin-Liang Chen}
\affiliation{Department of Physics and National Center for Theoretical Sciences, National
Cheng-Kung University, Tainan 701, Taiwan}
\author{Guang-Yin Chen}
\affiliation{Department of Physics, National Chung Hsing University, Taichung 402, Taiwan}
\author{Yueh-Nan Chen}
\email{yuehnan@mail.ncku.edu.tw}
\affiliation{Department of Physics and National Center for Theoretical Sciences, National
Cheng-Kung University, Tainan 701, Taiwan}
\date{\today }

\begin{abstract}
Entanglement plays a central role in the field of quantum information
science. It is well known that the degree of entanglement cannot be
increased under local operations. Here, we show that the concurrence of a
bipartite entangled state can be increased under the local $\mathcal{PT}$%
-symmetric operation. This violates the property of entanglement
monotonicity. We also use the Bell-CHSH and steering inequalities to
explore this phenomenon.
\end{abstract}

\pacs{03.67.Bg, 03.65.Ca, 03.67.Ac, 03.67.Hk}
\maketitle

\section{INTRODUCTION}

In conventional quantum mechanics, one of the axioms is that the Hamiltonian
of a closed system has to be Hermitian, leading to the properties of (1) The
eigenvalues of the Hamiltonian are real, and (2) the time-evolution of the
system is unitary. In 1998, Bender \emph{et~al.}~\cite{Bender} found that
the parity-time~($\mathcal{PT}$)-symmetric Hamiltonian, which is
non-Hermitian, can still have real energy spectra under some conditions.
Later, they~reconstructed the mathematical form of the inner product by
introducing \textit{C}-symmetry, such that the evolution of the $\mathcal{PT}
$-symmetric system becomes unitary \cite{Bender2}. In the Schr\"{o}%
dinger equation, the necessary but not sufficient condition for a
Hamiltonian to be $\mathcal{PT}$-symmetric is $V(x)=V^{\ast }(-x)$ \cite%
{Bender4}. Recently, experimental realizations of the $\mathcal{PT}$%
-symmetric Hamiltonian in classical optical systems have been proposed and
realized by using the spatially balanced gain and loss of energy \cite%
{Ganainy,Makris,Makris2,PT_exp,PT_exp2,PT_exp3,PT_exp4,PT_exp5,PT_exp6,PT_exp7}%
. However, even with the experimental success in classical optical systems, there
are still controversial results in some $\mathcal{PT}$-symmetric quantum systems. For
example, Bender \textit{et al}.~\cite{Bender3} found that the evolution
time between two quantum states under the $\mathcal{PT}$-symmetry operation can be
arbitrary small. Lee \textit{et~al}.~\cite{Lee} found that the no-signalling
principle can be violated when applying the local $\mathcal{PT}$-symmetric operation
on one of the entangled particles.

Quantum entanglement~\cite{entanglement} is one of the most intriguing
phenomena in quantum physics. Its history can be traced back to the
challenge by Einstein, Podolsky, and Rosen (EPR)~\cite{EPR}. In 1964, J.
Bell~\cite{Bell} proposed the famous ``Bell's inequality'' based on the
local hidden variable (LHV) model. Subsequent experiments~\cite{Bell_exp}
have successfully demonstrated violations of Bell's inequality,
meaning that quantum mechanics and the LHV theory are incompatible. In
response to the EPR paradox, Schr\"{o}dinger introduced a concept called
``quantum steering''. Steering was recently formalized
as a quantum information task by Wiseman \textit{et al.}~\cite{Wiseman}. The
steering inequality was further introduced \cite{SI} to delineate the
quantum steering from other non-local properties. For many years, Bell's
inequality has been used as an experimental tool~\cite{CHSH} to examine the
non-locality. Its relation with the steering inequality and entanglement has
also attracted great attention very recently \cite{Wiseman,GYChen}.

Motivated by these works, in this paper, we consider a bipartite system in which
one of the particles undergoes a local $\mathcal{PT}$-symmetric operation. We examine the behavior of the
bipartite entanglement through the concurrence, Bell's inequality, and the
steering inequality. Not only is the behavior of entanglement restoration
observed, but it is also found that its value can exceed the initial one. This violates the
property of entanglement monotonicity \cite{Bennett,Bennett2} and is beyond
the description of non-Markovian dynamics. We also show that the
increase of entanglement is not a unique property of the $\mathcal{PT}$%
-symmetric system by considering the non-Hermitian Hamiltonian without $%
\mathcal{PT}$-symmetry.

\section{RESTORATION OF ENTANGLEMENT BY LOCAL $\mathcal{PT}$-SYMMETRIC
OPERATION}

\begin{figure}[th]
\emph{\includegraphics[width=6cm]{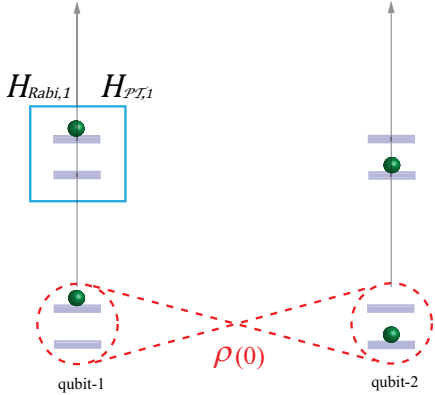} }
\caption{(Color online) The schematic diagram of the dynamics of the two
qubits. Qubit-1 undergoes either the Rabi process or a local $\mathcal{PT}$-symmetric
operation, while qubit-2 remains isolated. The initial condition is the
maximally entangled state: $(|00\rangle +|11\rangle )/\protect\sqrt{2}$.}
\label{fig01}
\end{figure}

As shown in Fig.~\ref{fig01}, we consider a composite system consisting of
two identical qubits. Let qubit-1 undergo a coherent Rabi oscillation
governed by $H_{\text{Rabi},1}=\hbar g\left( \sigma _{1,+}+\sigma
_{1,-}\right) $, where $\sigma _{1,+}$~($\sigma _{1,-}$) is the
raising~(lowering) operator and $\hbar g$ is the coupling strength. The
evolution of the entire system can be obtained by solving the following
equation
\begin{equation}
\dot{\rho}=\frac{1}{i\hbar }\left[ H,\rho (t)\right] ,  \label{master}
\end{equation}%
where $H=H_{\text{Rabi},1}\otimes I_{2}$ is the total Hamiltonian of the
composite system with $I_{2}$ denoting the identity operator of qubit-2.

Let us also consider a different scenario by replacing the coherent Rabi
process with a local $\mathcal{PT}$-symmetric operation on qubit-1. The
total Hamiltonian $H_{\mathcal{PT}}$ can then be written as~\cite{PT_Hami}
\begin{equation}
H_{\mathcal{PT}}=H_{\mathcal{PT},1}\otimes I_{2}=s%
\begin{pmatrix}
i\sin {\alpha } & 1 \\
1 & -i\sin {\alpha }%
\end{pmatrix}%
\otimes I_{2},  \label{HPT}
\end{equation}%
where $H_{\mathcal{PT},1}$ is the Hamiltonian of qubit-1. The real number $s$
is a scaling constant, and the real number $\alpha $ is the non-Hermiticity
of $H_{\mathcal{PT},1}$. The condition $|\alpha |<\pi /2$ keeps the
eigenvalues of $H_{\mathcal{PT},1}$ real, i.e., $\mathcal{PT}$ symmetric. The
non-Hermitian Hamiltonian $H_{\mathcal{PT},1}$ can be decomposed into a
Hermitian part ($H_{+}$) and an anti-Hermitian part ($H_{-}$)
\begin{equation}
\begin{split}
H_{\mathcal{PT}}& =s%
\begin{pmatrix}
0 & 1 \\
1 & 0%
\end{pmatrix}%
\otimes I_{2}+s%
\begin{pmatrix}
i\sin {\alpha } & 0 \\
0 & -i\sin {\alpha }%
\end{pmatrix}%
\otimes I_{2} \\
& =H_{+}+H_{-}.
\end{split}%
\end{equation}%
To obtain the evolution of the system, Eq.~(\ref{master}) has to be modified
as~\cite{IJMP13Sergi,PRL12Brody}
\begin{equation}
\dot{\rho}=\frac{1}{i\hbar }\left[ H_{+},\rho (t)\right] +\frac{1}{i\hbar }%
\{H_{-},\rho (t)\}.  \label{masterPT}
\end{equation}%
The solution can also be obtained by introducing a time-evolving
operator~\cite{Bender3,Lee}
\begin{equation}
U_{1}(t)=e^{-iH_{\mathcal{PT},1}t}=\frac{1}{\cos \alpha }%
\begin{pmatrix}
\cos (t^{\prime }-\alpha ) & -i\sin t^{\prime } \\
-i\sin t^{\prime } & \cos (t^{\prime }+\alpha )%
\end{pmatrix}%
,  \label{unitary}
\end{equation}%
where $t^{\prime }=\Delta E\cdot t$, with $\Delta E\equiv \frac{E_{+}-E_{-}}{%
2}$. Here, $E_{\pm }=\pm s\cos \alpha $ are the eigenvalues of $H_{\mathcal{%
PT},1}$.

\begin{figure*}[th]
\emph{\includegraphics[width=16cm]{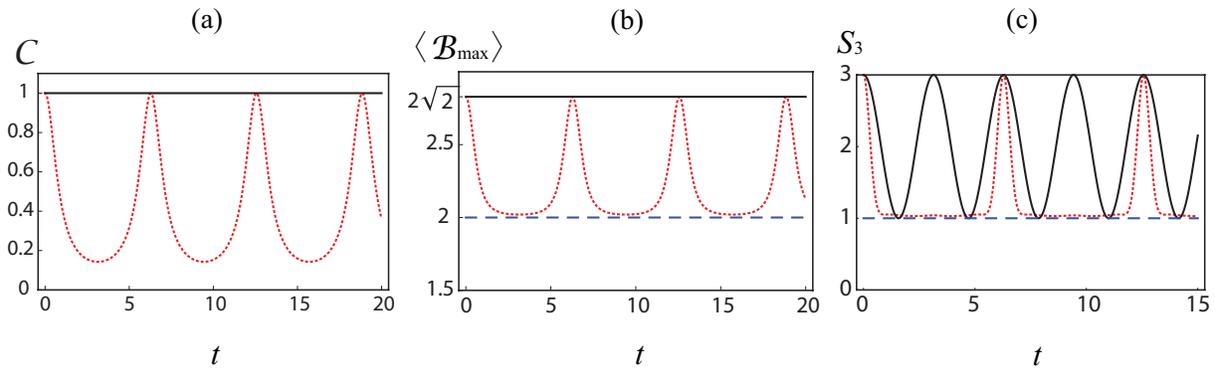} }
\caption{(Color online) The dynamics of (a)~the concurrence $\mathcal{C}$,
(b)~the maximal mean value of the Bell-kernel $\langle \mathcal{B}_{\text{max}%
}\rangle $, and (c)~the steering parameter $S_{3}$. The black-solid curve in
each figure represents that qubit-1 undergoes the coherent Rabi oscillation,
while the red-dotted curve represents that the local $\mathcal{PT}$%
-symmetric operation is performed on qubit-1. The blue-dashed lines in (b)
and (c) are the classical bounds (i.e., upper bounds in Eqs.~(\ref{BellI})
and (\ref{SteeringI}))
of $\langle \mathcal{B}_{\text{max}}\rangle
$ and $S_{3}$, respectively. In plotting the figure, the time $t$ is in units
of $1/\Delta E$, and the initial condition is $(|00\rangle +|11\rangle )/%
\protect\sqrt{2}$. }
\label{fig02}
\end{figure*}

In general, the evolution of a system with a non-Hermitian Hamiltonian is
not trace-preserving:
\begin{equation}
\frac{\partial }{\partial t}\text{tr}(\rho )=\frac{2}{i\hbar }\text{tr}%
\left( \rho H_{-}\right) \neq 0.
\end{equation}%
Thus, irrespective of whether one uses Eq.~(\ref{masterPT}) or Eq.~(\ref{unitary}), to
obtain a solution, $\rho (t)$ has to be renormalized,
\begin{equation}
\Tilde{\rho}(t)=\frac{\rho (t)}{\text{tr}(\rho (t))}  \label{renorm1}
\end{equation}%
or
\begin{equation}
\Tilde{\rho}(t)=\frac{\left( U_{1}(t)\otimes I_{2}\right) \rho (0)\left(
U_{1}(t)\otimes I_{2}\right) ^{\dagger }}{\text{Tr}(\left( U_{1}(t)\otimes
I_{2}\right) \rho (0)\left( U_{1}(t)\otimes I_{2}\right) ^{\dagger })},
\label{renorm2}
\end{equation}%
because the observers live in the conventional quantum world \cite%
{Lee,IJMP13Sergi,PRL12Brody}. The quantum average of an observable $\mathcal{%
A}$ can then be calculated as
\begin{equation}
\langle \mathcal{A}\rangle \equiv \text{tr}(\mathcal{A}\Tilde{\rho}(t))=%
\frac{\text{tr}(\mathcal{A}\rho (t))}{\text{tr}(\rho (t))}.
\label{qm_average}
\end{equation}%
In standard quantum mechanics, $\text{tr}(\rho (t))=1$, so Eq.~(\ref%
{qm_average}) coincides with the standard Born's rule.

To evaluate the degree of the entanglement between the two qubits, we use
the concurrence \cite{concurrence}
\begin{equation}
\mathcal{C(\rho )}=\text{Max}\left\{ 0,\sqrt{\lambda _{1}}-\sqrt{\lambda _{2}%
}-\sqrt{\lambda _{3}}-\sqrt{\lambda _{4}}\right\} ,
\end{equation}%
where $\left\{ \lambda _{i}\right\} $, in decreasing order, are the
eigenvalues of $\rho (\sigma _{y}\otimes \sigma _{y})\rho ^{\ast }(\sigma
_{y}\otimes \sigma _{y})$. Here, $\sigma _{y}$ is the Pauli-$y$ matrix, and $%
\rho ^{\ast }$ is the complex conjugate of $\rho $. To confirm the existence
of the entanglement experimentally, the Bell-CHSH inequality~\cite{Bell,CHSH}
and steering inequality~\cite{SI,steering_NC} are commonly used. Therefore,
it is useful to calculate the maximal mean-value of the Bell-kernel $\langle
\mathcal{B}_{\text{max}}\rangle $~\cite{Bell_Horodecki}
\begin{equation}
\langle \mathcal{B}_{\text{max}}\rangle =2\sqrt{M(\rho )}=2\sqrt{u_{1}+u_{2}}%
,
\end{equation}%
where $u_{1}$ and $u_{2}$ are the two largest eigenvalues of $T_{\rho
}^{T}T_{\rho }$, and $T_{\rho }^{T}$ is the transpose of $T_{\rho }$. The
correlation tensor $T_{\rho }$ is given by $t_{ij}=\text{Tr}\left[ \rho
(\sigma _{i}\otimes \sigma _{j})\right] $ for $i,j=1$, $2$, $3$, where $%
\sigma _{i}$ and $\sigma _{j}$ are the Pauli matrices.
If the correlation between the two qubits can be described by the LHV model,
the Bell-CHSH inequality holds~\cite{Bell_Horodecki}:
\begin{equation}
\langle \mathcal{B}_{\text{max}}\rangle \leq 2.
\label{BellI}
\end{equation}%
The violation of the Bell-CHSH inequality
indicates the failure of the LHV model and can be viewed as a
certification of quantum entanglement~\cite{Wiseman}.

If the correlation between two qubits can be described by the LHS model,
the steering inequality holds \cite{steering_NC}:
\begin{equation}
S_{N}\equiv \sum_{i=1}^{N}E\left[ \langle \hat{B_{i}}\rangle _{A_{i}}^{2}%
\right] \leq 1,
\label{SteeringI}
\end{equation}%
where $N(=2$ or $3)$ is the number of mutually unbiased measurement~\cite%
{MUB} (for example, the Pauli $\hat{X},\hat{Y}$ and $\hat{Z}$) performed on
qubit-2, and
\begin{equation}
E\left[ \langle \hat{B_{i}}\rangle _{A_{i}}^{2}\right] \equiv \sum_{a=\pm
1}P(A_{i}=a)\langle \hat{B_{i}}\rangle _{A_{i}=a}^{2}
\end{equation}%
is the average expectation of qubit-2. Here, $P(A_{i}=a)$ is the probability
of the measurement result of qubit-1, and
\begin{equation}
\langle \hat{B}\rangle _{A_{i}=a}\equiv \sum_{b=\pm 1}bP(B_{i}=b|A_{i}=a)
\end{equation}%
is the expectation value of qubit-2 conditioned on the outcome of qubit-1.
The violation of the steering inequality indicates the failure of the
LHS model and can also serve as an entanglement
certification~\cite{Wiseman}.

In Fig.~\ref{fig02}, we plot the dynamics of $\mathcal{C}$,
$S_{3}$, and
$\langle \mathcal{B}_{\text{max}}\rangle $ for both the coherent Rabi process
and the local $\mathcal{PT}$-symmetric operation. We set the initial state $%
|\psi _{AB}\rangle $ to be one of the Bell states: $|\psi _{AB}\rangle =%
\frac{1}{\sqrt{2}}(|00\rangle +|11\rangle )$. In Fig.~\ref{fig02}(a) and
(b), we can see that the values $\mathcal{C}$ and $\langle \mathcal{B}_{\text{max}%
}\rangle $ remain unchanged when qubit-1 undergoes the coherent Rabi
process. On the other hand, $\mathcal{C}$ and $\langle \mathcal{B}_{\text{max%
}}\rangle $ oscillate with time when performing the local $\mathcal{PT}$%
-symmetric operation. Thus, the entanglement between two parties can be
restored when one of them undergoes the $\mathcal{PT}$-symmetric operation.

\section{INCREASE OF ENTANGLEMENT BY LOCAL $\mathcal{PT}$-SYMMETRIC OPERATION}

\begin{figure}[th]
\emph{\includegraphics[width=6cm]{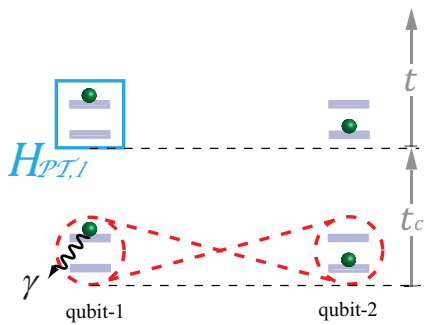} }
\caption{(Color online) The schematic diagram of the dynamics of the two
qubits. Qubit-1 is subjected to an amplitude damping for a time $t_{c}$. At
the time $t_{c}$, the damping is turned off, and the local $\mathcal{PT}$%
-symmetric operation is then performed on qubit-1. The initial state of the
system is $(|00\rangle +|11\rangle )/\protect\sqrt{2}$.}
\label{fig03}
\end{figure}

\begin{figure*}[th]
\emph{\includegraphics[width=16cm]{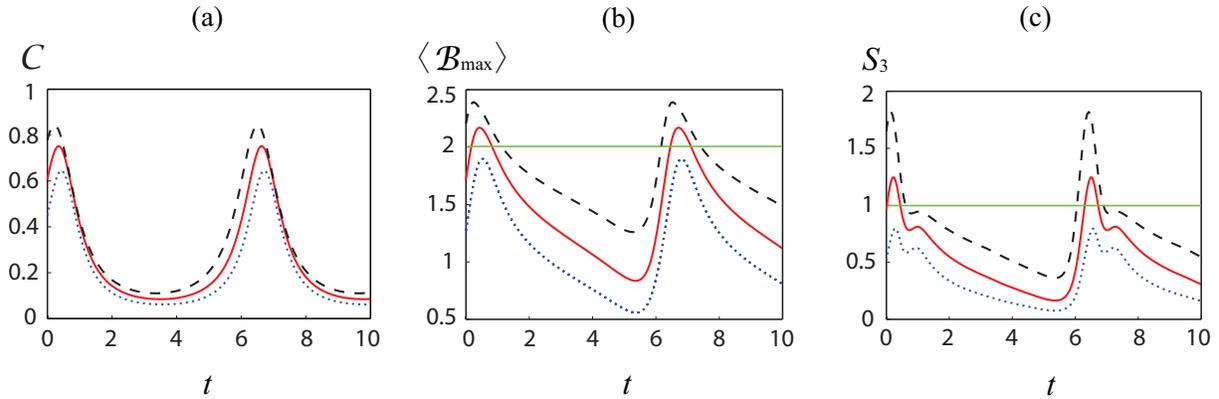} }
\caption{(Color online) The dynamics of (a)~the concurrence $\mathcal{C}$,
(b)~the maximal mean-value of the Bell-kernel $\langle \mathcal{B}_{\text{max}%
}\rangle $, and (c)~the steering parameter $S_{3}$ when the local $\mathcal{%
PT}$-symmetric operation is performed on qubit-1. The black-dashed,
red-solid, and blue-dotted curves represent the results of different initial
states of the $\mathcal{PT}$-symmetric evolution given by Eqs.~(\ref{IS1}), (%
\ref{IS2}), and (\ref{IS3}), respectively. The green horizontal lines in (b)
and (c) are the classical bounds (i.e., upper bounds in Eqs.~(\ref{BellI})
and (\ref{SteeringI}))
of $\langle \mathcal{B}_{\text{max}}\rangle
$ and $S_{3}$, respectively. In plotting the figure, the time $t$ of the $%
\mathcal{PT}$-symmetric evolution is in units of $1/\Delta E$.}
\label{fig04}
\end{figure*}

Let us consider qubit-1 embedded in an environment, while qubit-2 is
still isolated. From the angle of non-Markovian dynamics \cite{Rivas}, one
can observe the behavior of \textit{entanglement restoration} if there exists
some memory effect, i.e., quantum coherence is built between qubit-1 and the
environment during the evolution. So, one may speculate that the
entanglement restoration in Fig. 2 is similar to that derived from the non-Markovian
effect. However, we should note that the entanglement restored from the
environment cannot exceed the initial one for any non-Markovian process %
\cite{Rivas}. Otherwise, the property of entanglement
monotonicity is violated, i.e., entanglement cannot be created (or increased) by
performing any local operation~\cite{Bennett,Bennett2}. In this section, we
will show that the degree of the entanglement can exceed the initial value
for the local $\mathcal{PT}$-symmetric operation.

The first step is to prepare the quantum state which is not maximally
entangled. To accomplish this, let us start from the maximally-entangled
state, $\frac{1}{\sqrt{2}}(|00\rangle +|11\rangle )$, and subject qubit-1
to a Markovian amplitude damping (with rate $\gamma $) for a time $%
t_{c}$, as shown in Fig.~\ref{fig03}. By solving the following Lindblad-form~%
\cite{Lindblad,Lindblad2} master equation
\begin{equation}
\dot{\rho}_{c}=\frac{\gamma }{2}\left( 2\sigma _{1}^{-}\rho _{c}\sigma
_{1}^{+}-\sigma _{1}^{+}\sigma _{1}^{-}\rho _{c}-\rho _{c}\sigma
_{1}^{+}\sigma _{1}^{-}\right) ,
\end{equation}%
one can obtain the state $\rho _{c}(t_{c})$ at the cut-off time $t_{c}$:
\begin{equation}
\rho _{c}(t_{c})=\frac{1}{2}%
\begin{pmatrix}
e^{-\gamma t_{c}} & 0 & 0 & e^{-\gamma t_{c}/2} \\
0 & 0 & 0 & 0 \\
0 & 0 & 1-e^{-\gamma t_{c}} & 0 \\
e^{-\gamma t_{c}/2} & 0 & 0 & 1%
\end{pmatrix}%
\end{equation}%
with the concurrence $\mathcal{C=}e^{-\gamma t_{c}/2}$, which is a
monotonically decreasing function of $t_{c}$. For comparisons, we choose $%
t_{c}=0.5$, $1$, and $1.6$ (in units of $1/\gamma $) to have three different
initial states:
\begin{equation}
\rho _{c}(t_{c}=0.5)=%
\begin{pmatrix}
0.3033 & 0 & 0 & 0.3894 \\
0 & 0 & 0 & 0 \\
0 & 0 & 0.1967 & 0 \\
0.3894 & 0 & 0 & 0.5%
\end{pmatrix}%
,  \label{IS1}
\end{equation}%
\begin{equation}
\rho _{c}(t_{c}=1)=%
\begin{pmatrix}
0.1839 & 0 & 0 & 0.3033 \\
0 & 0 & 0 & 0 \\
0 & 0 & 0.3161 & 0 \\
0.3033 & 0 & 0 & 0.5%
\end{pmatrix}%
,  \label{IS2}
\end{equation}%
and
\begin{equation}
\rho _{c}(t_{c}=1.6)=%
\begin{pmatrix}
0.1009 & 0 & 0 & 0.2247 \\
0 & 0 & 0 & 0 \\
0 & 0 & 0.3991 & 0 \\
0.2247 & 0 & 0 & 0.5%
\end{pmatrix}%
.  \label{IS3}
\end{equation}%
With these three initial states, we let the system undergo the local $%
\mathcal{PT}$-symmetric operation (Eq.~(\ref{HPT})) to obtain the entanglement dynamics.
In Fig.~\ref{fig04}, we plot $\mathcal{C%
}$, $S_{3}$, and $%
\langle \mathcal{B}_{\text{max}}\rangle $ under the local $\mathcal{PT}$%
-symmetric operation. From $\langle \mathcal{B}_{\text{max}}\rangle $ (or $%
S_{3}$), we can see that it is possible to certify the entanglement (values
above the classical bound shown by the horizontal green line) at a later
time, even if initially the entanglement is not certified (values below the
classical bound). We can also see that the degree of the entanglement $%
\mathcal{C}$ can \textit{exceed} the initial value under the local $\mathcal{%
PT}$-symmetric operation. This violates the property of entanglement
monotonicity, i.e., entanglement cannot be created (or increased) by
performing any local operation~\cite{Bennett,Bennett2}.

To understand this thoroughly, let us examine the reduced density state of
qubit-2. In conventional quantum mechanics, a local operation on qubit-1
cannot alter the reduced state of qubit-2. Under the local $\mathcal{PT}$%
-symmetric operation, however, the reduced density states of qubit-2 at $%
t^{\prime }=0$ and $\pi /2$ are
\begin{equation}
\rho _{B}(0)=%
\begin{pmatrix}
\frac{1}{2} & 0 \\
0 & \frac{1}{2}%
\end{pmatrix}%
,
\end{equation}%
and
\begin{equation}
\rho _{B}(\frac{\pi }{2})=%
\begin{pmatrix}
\frac{1}{2} & \frac{i\sin \alpha }{(1+\sin ^{2}\alpha )} \\
\frac{-i\sin \alpha }{(1+\sin ^{2}\alpha )} & \frac{1}{2}%
\end{pmatrix}%
,
\end{equation}%
respectively, while the initial state is $\frac{1}{\sqrt{2}}(|00\rangle +|11\rangle )$.
This means that the reduced density state of qubit-2 is changed
under the local $\mathcal{PT}$-symmetric operation on qubit-1. From this
viewpoint, we know that the local $\mathcal{PT}$-symmetric operation is not
a genuine local operation in the conventional quantum world. More
importantly, when we deal with a $\mathcal{PT}$-symmetric Hamiltonian,
renormalization (Eqs.~(\ref{renorm1}) and~(\ref{renorm2})) is required to
ensure trace-preserving. Such an action directly affects the quantum
state of the two qubits and results in the violation of entanglement
monotonicity \cite{note}.

\begin{figure}[th]
\emph{\includegraphics[width=6.5cm]{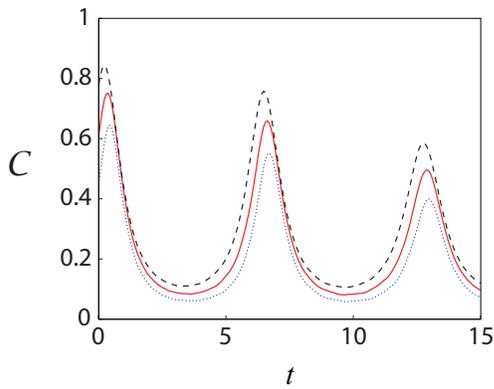} }
\caption{(Color online) The concurrence as a function of time when qubit-1
is under the Hamiltonian of Eq.~(\ref{breakingPT}). The black-dashed,
red-solid, and blue-dotted curves represent the results of the different initial
states given by Eqs.~(\ref{IS1}), (\ref{IS2}) and (\ref{IS3}) respectively.
In plotting the figure, the time $t$ is in units of $1/\Delta E$,
and $\protect\epsilon $ is set to 0.01. }
\label{fig05}
\end{figure}

To check whether the increase of entanglement is a unique property only for
the $\mathcal{PT}$-symmetric Hamiltonian, let us consider the following
Hamiltonian
\begin{equation}
H_{1}=s%
\begin{pmatrix}
i\sin {\alpha } & 1 \\
1 & -i\sin {\alpha }%
\end{pmatrix}%
+\epsilon
\begin{pmatrix}
1 & 0 \\
0 & -1%
\end{pmatrix}%
,  \label{breakingPT}
\end{equation}%
where $\epsilon $ is a real number. If $\epsilon $ is not equal to zero, $%
H_{1}$ is no longer a $\mathcal{PT}$-symmetric Hamiltonian but a normal
non-Hermitian Hamiltonian. As shown in Fig.~\ref{fig05}, the entanglement
can still be increased with the same initial conditions given in Eqs.~(\ref%
{IS1})-(\ref{IS3}). The reason is the same as that of the $\mathcal{PT}$%
-symmetric Hamiltonian. When we deal with the non-Hermitian Hamiltonian, the
renormalization procedure induces a non-local effect, and the entanglement
monotonicity is violated.

\section{CONCLUSION}

In summary, we find that the degree of entanglement between the two particles
oscillates with time, when the local $\mathcal{PT}$-symmetric operation is
performed on one of the qubits. To check whether this is similar to results
of non-Markovian effects, we consider a maximally-entangled state
subjected to Markovian damping for some time $t_{c}$, and then replace
the damping with the local $\mathcal{PT}$-symmetric operation. It is found
that the entanglement can be increased with the local $\mathcal{PT}$%
-symmetric operation. This contradicts the fact that entanglement cannot be
increased by any local operation. We also consider the non-Hermitian
Hamiltonian without $\mathcal{PT}$-symmetry, and show that the
increase of entanglement is not a unique property of the $\mathcal{PT}$%
-symmetric system.

\section{ACKNOWLEDGMENTS}

This work is supported partially by the National Center for Theoretical
Sciences and Ministry of Science and Technology, Taiwan, grant number MOST
101-2628-M-006-003-MY3, 102-2112-M-005-009-MY3 and 103-2112-M-006-017-MY4.


\begin{thebibliography}{99}
\bibitem{note} We should note that the reduced density matrix of qubit-2 is
changed even without the renormalization. However, the change of the reduced
state does not necessarily lead to the increase of the entanglement. It just
shows that the PT-symmetric operation is not a genuine local operation. For
example, when qubit-1 suffers Markovian amplitude damping (with a coherent
coupling between two qubits), the reduced state of qubit-2 is changed during
the evolution. Depending on the initial conditions, the entanglement may or
may not increase during the evolution.
\end{thebibliography}
\end{document}